\documentclass[5p,sort&compress]{elsarticle}
\usepackage[T1,T2A]{fontenc} 
\usepackage[cp1251]{inputenc}
\usepackage[english]{babel} 
\usepackage{subcaption} 
\usepackage{wrapfig}
\usepackage{graphicx}
\usepackage{epstopdf}
\usepackage{multicol}
\usepackage{tabularx}
\usepackage{amssymb,amsmath}
\usepackage{mathrsfs}
\usepackage{xfrac}
\usepackage{textcomp}
\usepackage{bm}
\usepackage{color}
\usepackage{ulem}
\usepackage{setspace} 
\usepackage{float}
\definecolor{dblue}{rgb}{0,0,0.8}

\definecolor{dred}{rgb}{0.7,0,0}

\definecolor{dgreen}{rgb}{0.,0.6,0}

\makeatletter 
\def\lambdabar{\protect\@lambdabar}
\def\@lambdabar{%
\relax
\bgroup
\def\@tempa{\hbox{\raise.73\ht0
\hbox to0pt{\kern.25\wd0\vrule width.5\wd0
height.1pt depth.1pt\hss}\box0}}%
\mathchoice{\setbox0\hbox{$\displaystyle\lambda$}\@tempa}%
{\setbox0\hbox{$\textstyle\lambda$}\@tempa}%
{\setbox0\hbox{$\scriptstyle\lambda$}\@tempa}%
{\setbox0\hbox{$\scriptscriptstyle\lambda$}\@tempa}%
\egroup
}
\makeatother

\usepackage{hyperref,lineno} %
\modulolinenumbers[5]
\journal{Physics Letters B}
\begin{document}

\begin{frontmatter}

\title{GEOMETRICAL OPTICS METHOD IN THE THEORY OF CHANNELING OF HIGH ENERGY PARTICLES IN CRYSTALS
}

\author[addr_KIPT,addr_KhUni]{N.F.~Shul'ga}\corref{mycorrespondingauthor}
\author[addr_KIPT,addr_KhUni]{S.N.~Shulga}
\cortext[mycorrespondingauthor]{Corresponding author, e-mail: shulga@kipt.kharkov.ua}
\address[addr_KIPT]{National Science Center "Kharkov Institute of Physics and Technology". 1, Akademichna str. Kharkiv 61108 Ukraine}
\address[addr_KhUni]{Karazin Kharkiv National University. 4, Svobody sq. Kharkiv 61000 Ukraine}

\begin{abstract}
The process of scattering of fast charged particles in thin crystals is considered in the transitional range of thicknesses, between those at which the channeling phenomenon is not developed and those at which it is realized. The possibility is shown of application of the methods of geometrical optics for description of the scattering process. The dependence is studied of the total scattering cross-section of ultrarelativistic positrons on target thickness in this range of crystal thicknesses. In the case of ultrarelativistic particles channeling the possibility is shown of the existence of an effect analogical to the Ramsauer-Townsend effect of conversion into zero of the total scattering cross-section at some values of crystal thickness. The important role is outlined of the Morse-Maslov index that enters into the wave function expression in the geometrical optics method.
\end{abstract}

\begin{keyword}
ultrathin crystal, scattering amplitude, scattering cross-section, channeling, operator method, geometrical optics.\\
\end{keyword}

\end{frontmatter}



\section{\label{sec:level1}Introduction}

At motion of a fast charged particle in crystal along one of crystal axes or planes the correlations appear between its consequent collisions with the lattice atoms. Due to these correlations, the channeling phenomenon is possible, at which the particles move in channels created by crystal atomic strings or atomic planes by periodically deviating from the channel direction to small angles (see \cite{Lindh1965, Gemmell1974, AkhiezShulga_HighEn1996} and references therein). At passing of particles through ultrathin crystals whose thickness is small compared to the distance of one channeling oscillation period, the channeling phenomenon is absent \cite{AkhiezShulga_HighEn1996}. In this case, there remains a possibility of manifestation of different coherent and interference effects in interaction of particles with lattice atoms. Some of these effects, connected with the study of angular distributions of the particles scattered by crystal were considered in the works \cite{ShuShu_PLB2017,ShuShuChe_NIMB2017} basing upon the operator method \cite{FeitFleck_JCP1982,Dabag1988,KozShuCherk2010} of the wave function determination in crystal. 

In the present work we demonstrate the possibility of application of the geometrical optics method (see \cite{Kravts_GeomOpt2011, Arnold_MathMeth1989}) in the problem of particle scattering in crystal that allows one to express the wave function and the quantum scattering cross-section by the means of the family of classical particle trajectories in crystal. This opens new possibilities for application of numerical methods to description of high energy particles scattering in the fields of complex configuration such as, for example, the field of crystal lattice, and of the analysis of quantum properties of the particle scattering process in this case. The main attention we pay to the study of the dependence of the total scattering cross-section on target thickness. The possibility is shown of the quantum effect of conversion into zero of the total scattering cross-section for some specific values of target thickness. This effect is analogical to the Ramsauer-Townsend effect \cite{Rams_AnnPh1921,Towns_PhMag1922} of conversion into zero of the total scattering cross-section of low energy electrons in the inert elements gases. In the considered problem, however, this effect is only revealed for ultrarelativistic particles. We indicate an important role in this problem of the Morse-Maslov index \cite{Kravts_GeomOpt2011, Arnold_MathMeth1989}, that enters into the expression of the wave function in the geometrical optics method.


\section{\label{sec:level2}Scattering amplitude and total scattering cross-section}

The total scattering cross-section of a relativistic charged particle in an external field $U\left( {\bf{r}} \right)$ is defined by the formula known as the optical theorem (see, e.g. \cite{AkhBer_QuantElectrod1965}):
\begin{equation}
\label{equ01}
{\sigma _{{\text{tot}}}} = \frac{{4\pi }}
{k}\operatorname{Im} a\left( {\bf{q}} \right)\left|
\begin{matrix}
   \hfill \cr 
  _{{\bf{q}} = 0} \hfill \cr 
\end{matrix}
\right.\,,
\end{equation}
where $\mathbf{p}$ is the particle momentum, $\mathbf{q}=\mathbf{p}-\mathbf{p}'$
 is the momentum transferred to the external field and $a\left( \mathbf{q}\right)$  
is the scattering amplitude
\begin{equation}
\label{equ02}
a\left( \mathbf{q}\right) =-\frac{i}{4\pi \hbar ^{2} } \int\nolimits_{V} d^{3} r\, \, e^{-\frac{i}{\hbar } \mathbf{p}' \mathbf{r}} \overline{u' } \gamma _{0} U\left( \mathbf{r}\right) \, \psi \left( \mathbf{r}\right).
\end{equation}
Here $\mathbf{p}'$ and $\overline{u'}$  are correspondingly the momentum and bispinor of the scattered particle and $\psi \left(\mathbf{r}\right)$ is the wave function of the initial particle.

With use of the Dirac equation for $\psi\left(\mathbf{r}\right)$ and of the Gauss theorem, the scattering amplitude \eqref{equ02} may be presented in the form of a surface integral taken around the area of the external field influence \cite{BonShu_PLB1998}:
\begin{equation}
\label{equ03}
a\left( \mathbf{q}\right) =-\frac{i}{4\pi \hbar } \oint\nolimits_{}^{}d\mathbf{S}\, \, \overline{u'} \bm\gamma\psi \left( \mathbf{r}\right) e^{-\, \frac{i}{\hbar } \mathbf{p}' \mathbf{r}}.
\end{equation}
This formula for the scattering amplitude is true for any arbitrary potential $U\left( \mathbf{r}\right)$. The only requirement imposed onto this potential is that it was localized in a spatial region of the volume $V$. 

Let us note that the form of an enclosed surface in the integral entering into \eqref{equ03} can be arbitrary. At solution of a number of problems it is convenient to choose a surface of a cylinder whose borders are perpendicular to the momentum $\mathbf{p}$ of incident particle and are situated near the region of entering of the particle into the external field and its exit from this field, and its lateral side is moved away to a large distance from the region where the potential is not zero (see Fig.~1). By neglecting this contribution into the amplitude \eqref{equ03} from the lateral side of the cylinder, we come to the following expression for the scattering amplitude:
\begin{figure}
\centering
\includegraphics[width=0.8\columnwidth]{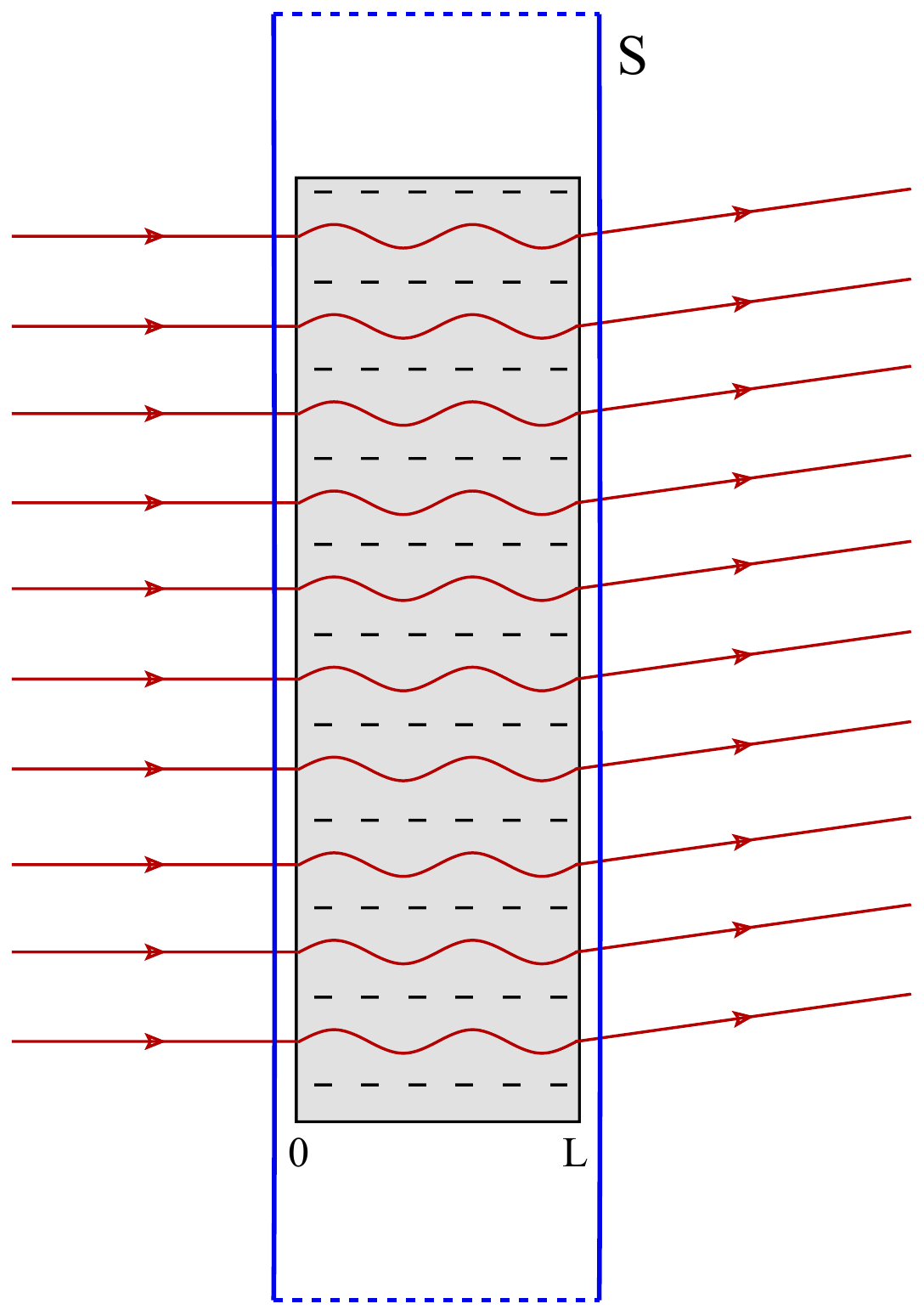}
\caption{\label{fig1}Crystal as a target for incident particles and the surface $S$ surrounding it.}
\end{figure}
\begin{equation}
\label{equ04}
a\left( \mathbf{q}\right) =-\frac{i}{4\pi \hbar } \left. \int\nolimits_{}^{}d^{2} \rho \, \, e^{-\, \frac{i}{\hbar } \mathbf{p}' \mathbf{r}} \overline{u' } \gamma _{z} \, \psi \left( \mathbf{r}\right)  \right| _{z=0}^{z=L},
\end{equation}
where the $z$ axis is parallel to the incident particle momentum, $\bm\rho=\left( x,y\right)$  are the coordinates in the transversal plane and $L$ is the thickness of the target interacting with the particle.


\section{\label{sec:level3}Continuous potential of crystal axes and planes}

At deduction of the formula \eqref{equ04} the concrete form of the potential $U\left( \mathbf{r}\right)$ was not used, therefore this formula can be used for calculation of the scattering amplitude in the fields of quite complex configuration, such, for example, as the field of a totality of crystal lattice atoms
\begin{equation}
\label{equ05}
U\left( \mathbf{r}\right) =\sum\limits_{n=0}^{N}u\left( \mathbf{r}-\mathbf{r}_{n} \right),
\end{equation}
where $u\left( \mathbf{r}-\mathbf{r}_{n} \right)$  is the potential of the field of each separate lattice atom with the position $\mathbf{r}_{n}$. 

A particular interest represents the case of incidence of fast charged particles onto a thin crystal along one of crystallographic axes or planes. This case is interesting because at such motion there appear correlations between consequent collisions of a particle with the lattice atoms, that can result in the phenomenon of channeling of particles in the crystal, at which the particles move inside the channels formed by crystal atomic strings or crystal atomic planes periodically situated in the crystal. The substantial thing here is that at passing of high energy particles through thin crystals one can neglect the influence of incoherent effects onto the motion, that are connected with the thermal spread of the atoms positions in the lattice. In this case the particles motion will be determined basically by the continuous potential pf crystal atomic strings or planes, depending on the conditions of entering of particles into the crystal \cite{Lindh1965, Gemmell1974, AkhiezShulga_HighEn1996}. The continuous potential of crystal atomic strings is the lattice potential averaged along the string axis:
\begin{equation}
\label{equ06}
U_{0} \left(\bm\rho\right) =\frac{1}{L} \int\nolimits_{0}^{L}dz\, U\left( \bm\rho,z\right), \qquad 0\leq z\leq L.
\end{equation}
The continuous potential of crystal atomic planes is the potential $U\left(\mathbf{r}\right)$ averaged over the coordinates $\left( y,z\right)$:
\begin{equation}
\label{equ07}
U_{pl} \left(x\right) =\frac{1}{L_{y} L} \int\nolimits dy\, dz\, U\left( x,y,z\right), \qquad 0\leq z\leq L,
\end{equation}
where $L$ and $L_y$ are the crystal thicknesses along the coordinates $z$ and $y$. 

The continuous potential approximation allows one to significantly facilitate the description of the particle motion in crystal in the frames of both classical and quantum theories, by reducing the problem to the motion of a particle in a two-dimension \eqref{equ06} and one-dimension \eqref{equ07} potentials. 


\section{\label{sec:level4}Coherent scattering}

For calculation of the wave function of an electron moving in the fields \eqref{equ06} or \eqref{equ07} we will take profit of the squared Dirac equation with the initial condition that represents entering of a plane wave into the crystal. By neglecting in this equation the spin-field interaction and ${U_{c} }/{\varepsilon}$, we come to the following equation for the function $\varphi \left( \mathbf{r}\right)$, that determines in the external field the wave function deviation $\psi \left( \mathbf{r}\right) =\varphi \left( \mathbf{r}\right) u_{p}\,e^{{i\mathbf{p}\mathbf{r}}/{\hbar } }$ from the plane wave \cite{ShuShu_PLB2017}:
\begin{equation}
\label{equ08}
i\hbar v{\partial _z}\varphi  = \left[ {\frac{{{\bf{\hat p}}_ \bot ^2}}
{{2\varepsilon }} + U\left( {{\bm{\rho }},z} \right)} \right]\varphi,
\end{equation}
where ${\bf{\hat p}}_\bot = - i\hbar \left(\partial / \partial{\bm{\rho}} \right)$, ${\bf{p}}$ and ${u_p}$ are correspondingly the momentum and bispinor of the plane wave incident onto the crystal.

Despite the simplifications made, the solution of the equation \eqref{equ08} represents itself quite a hard problem in view of a complex form of the function $U_{c} \left(\bm\rho\right)$ (see, e.g. \cite{AkhiezShulga_HighEn1996}). Therefore for determination of the function $\varphi\left(\bm\rho,z\right)$, satisfying the condition $\varphi \left(\bm\rho,z\right)\left|\begin{array}{l}
 \\
_{z\leq 0} 
\end{array}
\right. =1 $, the use of approximated or numerical methods is required.

In the simplest case of high energies and ultrasmall crystal thicknesses, where the particle motion is similar to rectilinear, the eikonal approximation is applicable for obtaining the function $\varphi \left( \mathbf{r}\right)$. In this approximation the function $\varphi \left( \mathbf{r}\right)$ has the following form \cite{AkhiezShulga_HighEn1996,Newt66}:
\begin{equation}
\label{equ09}
{\varphi_{_{eik}}}({\bf{r}}) = \exp \left\{ {\frac{i}
{\hbar }\int\limits_0^z {dz'{U_c}\left( {{\bm{\rho }},z'} \right)} } \right\}.
\end{equation}

By using this function, we come to the following expression for the scattering amplitude in the eikonal approximation:

\begin{equation}
\label{equ10}
{a^{eik}}\left( {{{\bf{q}}_ \bot }} \right) = \frac{{i\,p}}
{{2\pi \hbar }}\int {{d^2}\rho \:{e^{i{\bf{q}\bm{\rho}}/\hbar }}\left( {1 - {e^{\frac{i}
{\hbar }{\chi _0}\left( {{\bm{\rho }},\,L} \right)}}} \right)},
\end{equation}
where $L$ is the crystal thickness and  
$$
{\chi _0}\left( {{\bm{\rho }},\,L} \right) =  - \frac{L}
{v}{U_c}\left( {\bm{\rho }} \right).
$$

At $L \to 0$ the formula \eqref{equ10} is transformed into the corresponding result of the Born approximation, according to which the scattering amplitude is proportional to the target thickness:
\begin{equation}
\label{equ11}
{a^B}\left( {{{\bf{q}}_ \bot }} \right) = \frac{\varepsilon }
{{2\pi {\hbar ^2}}}L\int {{d^2}\rho \:{e^{i{\bf{q}\bm\rho}/\hbar }}{U_c}\left( {\bm{\rho }} \right)}.
\end{equation}
In this case the differential and total scattering cross-sections are proportional to the square of crystal thickness (or to the square of the number of atoms with which the particle interacts while passing through crystal). This effect is known as the coherent scattering process at particles scattering in crystal \cite{AkhiezShulga_HighEn1996}. This effect is possible if $\left| {{\chi _0}} \right| \ll \hbar $
, that corresponds by the order of value to satisfying the condition 
\begin{equation}
\label{equ12}
N\frac{{Z{e^2}}}{{\hbar v}} \ll 1,
\end{equation}
where $N = L/a$, $a$ is the distance between crystal atoms, with which the particle interacts, and $Z\left| e \right|$ is the charge of a separate crystal atom.

With the target thickness increase, at violation of the condition \eqref{equ12}, the coherent effect in scattering destroys. In this case the quadratic dependence of both differential and total scattering cross-sections on $L$ are substituted by a more weak logarithmic dependence \cite{AkhiezShulga_HighEn1996,Kalash_JETPL1972}. 

The analysis of corrections to the eikonal approximation in the considered problem shows that the applicability condition of this approximation in the considered problem \cite{AkhiezShulga_HighEn1996} 
\begin{equation}
\label{equ13}
\frac{{{N^2}a}}{{p{R^2}}}Z{e^2} \ll 1
\end{equation}
quickly destroys with the target thickness increase, where $R$ is the screening radius of the separated crystal atom potential. This inequality, in fact, represents itself the condition of a weak deflection of particle trajectory in crystal from rectilinear motion.  At violation of this condition, the phenomena become possible of channeling and above-barrier motion of particles in the field of continuous potential of crystal atomic strings \eqref{equ06} and atomic planes \eqref{equ07}. In this case, the analysis of the particle scattering process in crystal requires a special consideration that allows to go aver the frames of the eikonal approximation.

\section{\label{sec:level5}Geometrical optics approximation}

The description of the scattering process of a particle by crystal at violation of the condition \eqref{equ13} can be performed basing upon the geometrical optics approximation. This approximation represents one of the variants of the quasiclassical approximation of quantum electrodynamics in which the wave function is determined by a family of classical particle trajectories in an external field \cite{Kravts_GeomOpt2011, Arnold_MathMeth1989}. 

In the quasiclassical approximation, the wave function in the field $U\left( {\bf{r}} \right)$ has the following form \cite{AkhiezShulga_HighEn1996, Arnold_MathMeth1989}:
\begin{equation}
\label{equ14}
\psi^{WKB} \left(\bm\rho,z\right) =\sqrt{f\left(\bm\rho,z\right) } \, u\, e^{\frac{i}{\hbar } S\left(\bm\rho,z\right) },
\end{equation}
where $S\left( \mathbf{r}\right)$ is the classical action satisfying the Hamilton-Jacobi equation
\begin{equation}
\label{equ15}
\left( \varepsilon -U\left( \mathbf{r}\right) \right) ^{2} =\left( \nabla S\right) ^{2} +m^{2}.
\end{equation}
The pre-exponent function $f$ in \eqref{equ14} is a function satisfying the continuity condition%
\begin{equation}
\label{equ16}
\nabla S\cdot \nabla f+\left( \nabla ^{2} S\right) f=0.
\end{equation}
The bispinor $u$ entering into \eqref{equ14}, in neglecting the interaction spin-field, represents the bispinor $u_{0}$ of the incident particle. At high energies, in the action $S\left( \mathbf{r}\right) =\mathbf{p}\mathbf{r}+\chi \left( \mathbf{r}\right)$ it is convenient to separate the term $\mathbf{p}\mathbf{r}$ determining the plane wave part of the wave function of the incident particle. Then the function $\chi \left( \mathbf{r}\right)$ will define the addition to the wave function phase \eqref{equ14} that is connected with the interaction of the particle with the external field. This function, according to \eqref{equ15}, will be determined in the approximation $\left| U\left( \mathbf{r}\right) \right| \ll \varepsilon$ by the following equation:
\begin{equation}
\label{equ17}
- v{\partial _z}\chi  = U\left( {{\bm{\rho }},z} \right) + \frac{1}
{{2\varepsilon }}{\left( {{\nabla _ \bot }\,\chi \left( {{\bm{\rho }},z} \right)} \right)^2}.
\end{equation}
The solution of this equation in the area of the external field action $0\leq z\leq L$ has the following form:
\begin{equation}
\label{equ18}
\chi \left(\bm\rho\left( \mathbf{b},z\right),z\right) =
-\frac{1}{v}
\int_{0}^{z}dz'
\left[ 2U_{c} \left(\bm\rho\left( \mathbf{b},z' \right) \right) -U_{c} \left( \mathbf{b}\right) \right],
\end{equation}
where $\mathbf{b}$ is the value of the transversal coordinate at the particle entering into the crystal and $\bm\rho\left(\mathbf{b},z\right)$ is the trajectory of the transversal motion of a particle that is the solution of the equation of motion \cite{Lindh1965,Gemmell1974,AkhiezShulga_HighEn1996}
\begin{equation}
\label{equ19}
\frac{d^{2} }{dz^2}\bm\rho =-\frac{c^{2} }{\varepsilon ^{2} } \frac{\partial }{\partial\bm\rho } U_{c} \left(\bm\rho\right)
\end{equation}
with the initial condition $\bm\rho\left(\mathbf{b},z\right) \left| 
\begin{array}{l} \\_{z=0} \end{array} \right. =\mathbf{b} $.

The solution of the continuity equation \eqref{equ16} is determined by the following determinant \cite{Kravts_GeomOpt2011, Arnold_MathMeth1989}:
\begin{equation}
\label{equ20}
D=\frac{\partial \left( x,y,z\right) }{\partial \left( b_{x},b_{y},\tau \right) } =\det \left( 
\begin{array}{ccc}
\partial _{b_{x} } x & \partial _{b_{y} } x & \partial _{\tau } x \\
\partial _{b_{x} } y & \partial _{b_{y} } y & \partial _{\tau } y \\
\partial _{b_{x} } z & \partial _{b_{y} } z & \partial _{\tau } z
\end{array}
\right),
\end{equation}
where $\tau$ is the length of the particle trajectory in crystal (in our case $\tau \approx z $ ). This solution determines the connection of a family of particle trajectories in the depth $z$ of their penetration into the crystal, with the values $\mathbf{b}$ of initial coordinates of the wave entering into the crystal.

By using the obtained expressions for $\chi$ and $D$, we come to the following expression for the wave function $\psi \left(\bm\rho,L\right)$ that enters into the scattering amplitude \eqref{equ04}  
\begin{equation}
\label{equ21}
\psi \left(\bm\rho\left( \mathbf{b},L\right),L\right) =\frac{1}{\sqrt{\left| D\right| } } \exp \left\{ ipL+i\chi \left(\bm\rho\left( \mathbf{b},L\right),L\right) -i\frac{\pi }{2} \mu \right\},
\end{equation}
where $\mu$ is the Morse-Maslov index \cite{Kravts_GeomOpt2011, Arnold_MathMeth1989} that is connected with the number of transitions of the determinant $D$ through zero on the interval $0<z<L$.

\section{\label{sec:level6}Total scattering cross-section in the field of crystal atomic planes}

Let us consider the simplest variant of the formulas obtained above, in which an ultrarelativistsic positron enters into the crystal along crystal atomic planes. The continuous interplanar potential \eqref{equ07} in this case, with a good precision, may be approximated by a parabolic function \cite{Gemmell1974,AkhiezShulga_HighEn1996}
\begin{equation}
\label{equ22}
U_{} \left( x\right) =U_{0} \frac{x^{2} }{\left( a/2\right) ^{2} },
\qquad\quad
\left| x\right| \leq {a}/{2}
\end{equation}
(in the case of the motion in the field of crystal planes we will designate as $a$ the distance between the adjacent crystal atomic planes). For the Si $\left( 110\right)$ atomic planes, for example, $U_{0} \approx 22.7\, \, \text{eV}$. 

\begin{figure*}
\centering
\includegraphics[width=0.95\textwidth]{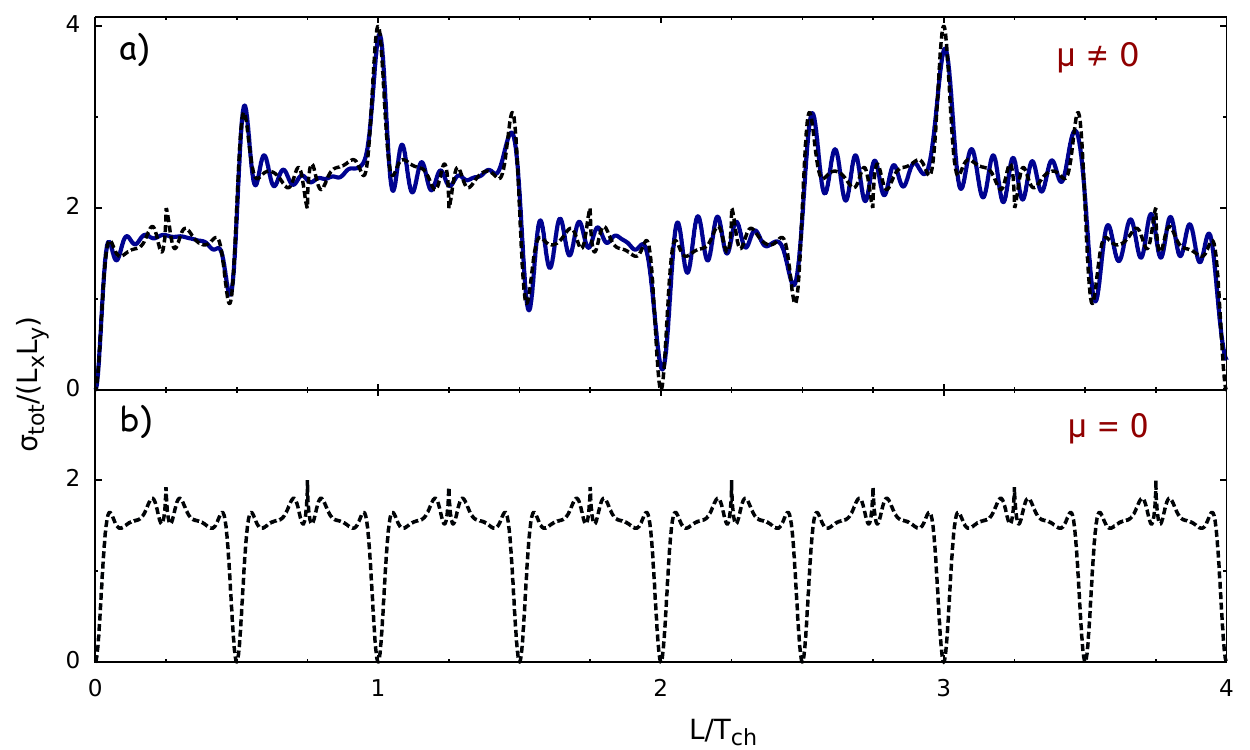}
\caption{\label{fig2}Total scattering cross-section calculated from the wave function obtained with use of the operator method (whole line) and its quasiclassical estimations (dashed lines) with the Morse-Maslov index taken not equal to zero (above) and equal to zero (below).}
\end{figure*}

At motion in such a field, according to \eqref{equ19}, a particle performs a harmonic motion in the transversal direction
\begin{equation}
\label{equ23}
x\left( z\right) =b\cos \Omega z
\end{equation}
with the frequency $\Omega=\sqrt{{8U_{0} }/{\varepsilon a^{2} } }$. For such a motion the functions $\chi \left( \mathbf{r}\right)$ and $D\left( \mathbf{r}\right)$ are calculated analytically: 
\begin{equation}
\label{equ24}
\chi \left( \mathbf{r}\right) =-\frac{b^{2} }{\left( {a}/{2} \right) ^{2} } \frac{U_{0} }{2v\Omega } \sin 2\Omega z, \qquad D\left( \mathbf{r}\right) =\cos \Omega z.
\end{equation}
The total scattering cross-section \eqref{equ01} in this case takes the following form:
\begin{equation}
\label{equ25}
\begin{array}{l}
\sigma_{t} =2L_{y} L_{x} \Big\{
   1-\mathop{\rm Re} \left|\cos \Omega L\right|^{1/2} \times\\
      \quad\;\times\;\int_{-a/2}^{a/2}
         \frac{db}{a}
         \exp\left[
            -\frac{i}{\hbar v}\frac{2U_0}{\Omega}\frac{b^2}{a^2}
            \sin 2\Omega L-i\frac{\pi}{2}\mu
         \right]
   \Big\},
\end{array}
\end{equation}
where $L_{x}$ and $L_{y}$ are linear dimensions of a crystal along the $x$ and $y$ axes. In calculation of the zero angle scattering amplitude we switched over in \eqref{equ04} from the integration over the final coordinate $x$ to the integration over impact parameters $b$ with use of the relation $dx\left(b\right) =\left| {dx}/{db} \right|db$.

At $\varepsilon \rightarrow \infty$, we have $\Omega\rightarrow 0$. The value $\mu$ in this case is equal to zero. In this case the formula \eqref{equ25} transforms into the following result of the eikonal approximation \eqref{equ10}.

In Fig.~2 we present the dependence \eqref{equ25} of the total scattering cross-section on the target thickness $L$ for positrons with the energy $\varepsilon =100\, \text{MeV}$, incident onto the silicon crystal alongside crystal planes $\left(110\right)$. The ordinate axis is the value ${\sigma _{t} }/{\left( L_{x} L_{y} \right)}$, and the abscissa axis is the value ${L}/{T}$, where $T=\Omega/{2\pi}$ is the period of the positron oscillations inside the channel. On the same plot we present the results of calculation of the dependence of $\sigma_{t}$ on $L$ obtained basing upon the operator method of the wave function determination (for the details of calculation with use of this method we refer the reader to \cite{ShuShu_PLB2017}). 

The obtained results show that both methods bring to quite similar results of calculations. The dependence of the total scattering cross-section on target thickness, as the results show, is quite complex. Let us pay attention to some particularities of this dependence. 

At $L\rightarrow 0$, the coherent effect in scattering takes place, according to which the total scattering cross-section $\sigma _{t}$ is proportional to the square of the target thickness, $\sigma_{t} \propto L^{2}$. With the target thickness increase this effect destroys, and the scattering cross-section is about a constant value. At the values of $L$ that are close to the half-period of oscillation of the channelled positron, the total scattering cross-section, as Fig.~2 shows, suffers substantial changes caused by the change of the Morse-Maslov index at each conversion to zero of the determinant \eqref{equ20} with the increase of $L$ and the structure of the determinant entering into \eqref{equ25}. 

Let us pay a particular attention to the fact that at the values of $L$ equal to the whole number values of the doubled period of the particle oscillations in crystal ($L\approx 2T$), the total scattering cross-section is close to zero. The uncommonness of this phenomenon consists in the fact that the positrons, while passing through crystal, are subject to the crystal lattice field impact. But the scattering of particles in absent. We outline that this effect takes place for ultrarelativistic particles. By the other words, the field and the force acting to the particle are present but the scattering is absent!

This effect is analogical to the Ramsauer-Townsend effect of conversion into zero of the total scattering cross-section of low energy electrons ($\varepsilon \approx 0.7\, \text{eV}  $ ) on the atoms of the inert elements gases (Ar, Kr, Xe,...), that was discovered experimentally at the dawn of the quantum mechanics creation \cite{Rams_AnnPh1921, Towns_PhMag1922, MottMas_ThAtCollis_1965}. Niels Bohr was delighted with this effect, so far as it is a direct manifestation of the quantum properties of matter. The particularity of the effect considered in the present work is that it takes place for ultrarelativistic particles. Let us note that, if we will formally suppose $\mu=0$ than the dependence of the total scattering cross-section on thickness changes substantially (see Fig.~2b). In this case, the conversion of $\sigma_t$ into zero takes place at each half-period of positron oscillations in the channel.

It could seem that the considered effect of conversion to zero of the total scattering cross-section would be characteristic for the positively charged particles only, for which the oscillation period is constant on entire impact parameters range. Nevertheless, the analysis of the dependence $\sigma _{t} =\sigma _{t} \left( L\right)$ for electrons at scattering in the continuous potential of crystal atomic planes \eqref{equ07} shows that an analogical effect is also possible for such particles. It is, however, expressed less brightly than for the positively charged particles. An analogical effect is also possible in a more complex case of scattering in a two-dimensional potential of crystal atomic strings \eqref{equ06}. Indicative in this connection is the evolution of the dependence of angular distributions of scattered particles with the crystal thickness increase (see Fig.~3 in \cite{ShuShu_PLB2017}). A detailed analysis of the process of particles scattering in a crystal in these cases, however, requires a particular consideration.


\section{References}

\bibliography{mybibfile}

\begin{thebibliography}{10}
\expandafter\ifx\csname url\endcsname\relax
  \def\url#1{\texttt{#1}}\fi
\expandafter\ifx\csname urlprefix\endcsname\relax\def\urlprefix{URL }\fi
\expandafter\ifx\csname href\endcsname\relax
  \def\href#1#2{#2} \def\path#1{#1}\fi

\bibitem{Lindh1965}
J.~Lindhard, Influence of crystal lattice on motion of energetic charged
  particles, Kon. Dansk. Vidensk. Mat.-Fys. Medd. 34 no.\,14 (1965) 1--64.

\bibitem{Gemmell1974}
D.~Gemmell, Channeling and related effects in the motion of charged particles
  through crystals, Rev. of Mod. Phys. 46 (1974) 129.
\newblock \href {http://dx.doi.org/10.1103/RevModPhys.46.129}
  {\path{doi:10.1103/RevModPhys.46.129}}.

\bibitem{AkhiezShulga_HighEn1996}
A.~Akhiezer, N.~Shul'ga, High Energy Electodynamics in Matter, Gordon and
  Breach, Amsterdam, 1996.

\bibitem{ShuShu_PLB2017}
N.~Shul'ga, S.~Shulga, Scattering of ultrarelativistic electrons in ultrathin
  crystals, Phys. Lett. B 769 (2017) 141--145.
\newblock \href {http://dx.doi.org/10.1016/j.physletb.2017.03.041}
  {\path{doi:10.1016/j.physletb.2017.03.041}}.

\bibitem{ShuShuChe_NIMB2017}
S.~Shulga, N.~Shul'ga, S.~Barsuk, I.~Chaikovska, R.~Chehab, On classical and
  quantum effects at scattering of ultrarelativistic electrons in ultrathin
  crystal, Nucl. Instr. Meth. B 402 (2017) 16--20.
\newblock \href {http://dx.doi.org/10.1016/j.nimb.2017.03.024}
  {\path{doi:10.1016/j.nimb.2017.03.024}}.

\bibitem{FeitFleck_JCP1982}
M.~Feit, J.~{Fleck Jr.}, A.~Steiger, Solution of the {Schr\"{o}dinger} equation
  by a spectral method, Journ. of Comp. Phys. 47 (1982) 412--433.
\newblock \href {http://dx.doi.org/10.1016/0021-9991(82)90091-2}
  {\path{doi:10.1016/0021-9991(82)90091-2}}.

\bibitem{Dabag1988}
S.~Dabagov, L.~Ognev, Passage of {MeV}-energy electrons through monocrystals,
  Nucl. Instr. Meth. B 30 (1988) 185--190.
\newblock \href {http://dx.doi.org/10.1016/0168-583X(88)90115-2}
  {\path{doi:10.1016/0168-583X(88)90115-2}}.

\bibitem{KozShuCherk2010}
A.~Kozlov, N.~Shul'ga, V.~Cherkaskiy, Spectral method in quantum theory of
  channeling phenomena of fast charged particles in crystals, Phys. Lett. A 374
  (2010) 4690--4694.
\newblock \href {http://dx.doi.org/10.1016/j.physleta.2010.09.025}
  {\path{doi:10.1016/j.physleta.2010.09.025}}.

\bibitem{Kravts_GeomOpt2011}
Y.~Kravtsov, Y.~Orlov, Geometrical optics of inhomogeneous media,
  Springer-Verlag, Berlin, 2011.

\bibitem{Arnold_MathMeth1989}
V.~Arnold, Mathematical methods in classical mechanics, Springer-Verlag, NY,
  1989.

\bibitem{Rams_AnnPh1921}
C.~Ramsauer, \"{U}ber den {W}irkungsquerschnitt {d}er {G}asmolek\"{u}le
  {g}egen\"{u}ber {l}angsamen {E}lektronen, Annalen der Physik 369 (1921)
  513--540.
\newblock \href {http://dx.doi.org/doi.wiley.com/10.1002/andp.19213690603}
  {\path{doi:doi.wiley.com/10.1002/andp.19213690603}}.

\bibitem{Towns_PhMag1922}
J.~Townsend, V.~Bailey, The motion of electrons in argon, Philosophical
  Magazine 43 (1922) 593--600.

\bibitem{AkhBer_QuantElectrod1965}
A.~Akhiezer, V.~Berestetskij, Quantum Electrodynamics, Interscience, NY, 1965.

\bibitem{BonShu_PLB1998}
N.~Bondarenko, N.~Shul'ga, The {Gauss} theorem in potential scattering theory
  and semi-classical corrections to the eikonal scattering amplitude, Phys.
  Lett. B 427 (1998) 114--118.
\newblock \href {http://dx.doi.org/10.1016/S0370-2693(98)00244-5}
  {\path{doi:10.1016/S0370-2693(98)00244-5}}.

\bibitem{Newt66}
R.~Newton, Scattering Theory of Waves and Particles, McGraw-Hill Book Company,
  NY, 1966.

\bibitem{Kalash_JETPL1972}
N.~Kalashnikov, E.~Koptelov, M.~Ryazanov, Destructive interference in the
  scattering of fast electrons in a single crystal, JETP Letters 15 (1972)
  82--85.

\bibitem{MottMas_ThAtCollis_1965}
N.~Mott, H.~Massey, The theory of atomic collisions, CP, Oxford, 1965.

\end{thebibliography}

\end{document}